\documentclass[structabstract]{aa}

\usepackage{graphicx}
\usepackage{amssymb}

\usepackage{natbib}

\usepackage{txfonts}

\begin{document}

\title{Testing Dark Energy models vs $\Lambda$CDM Cosmology by Supernovae and Gamma Ray Bursts}
\titlerunning{Equation of State by Gamma Ray Bursts}
\authorrunning{L. Izzo. et al}
\author{L. Izzo\inst{1,2,3}, S. Capozziello\inst{1}, G. Covone\inst{1}
\and M. Capaccioli\inst{1,4}}

\institute{Dipartimento di Scienze Fisiche, Universit\`a di Napoli
"Federico II" and INFN Sez. di Napoli, Compl. Univ. Monte S.
Angelo, Ed. N, Via Cinthia, I-80126 Napoli, Italy, \and ICRANet
and ICRA, Piazzale della Repubblica 10, I-65122 Pescara, Italy,
\and Dip. di Fisica, Universit\`a di Roma "La Sapienza", Piazzale
Aldo Moro 5, I-00185 Roma, Italy,  \and INAF - VSTceN, Salita
Moiariello, 16, I-80131, Napoli, Italy}

\abstract {}{A new method to constrain the cosmological equation
of state is proposed by using combined samples of gamma-ray bursts
(GRBs) and supernovae (SNeIa). }{The Chevallier-Polarski-Linder
parameterization is adopted for the equation of state in order to
find out  a realistic approach to achieve the
deceleration/acceleration transition phase of dark energy
models.}{As results, we find that GRBs, calibrated by SNeIa, could
be, at least,  good distance indicators capable of discriminating
cosmological models with respect to $\Lambda$CDM at high redshift.
Besides, GRBs+SNeIa combined redshift-distance diagram puts better
in evidence the change of slope around redshift $z\sim 0.5$ which
is usually addressed as the "signature" of today observed
acceleration. This feature could be interpreted, in more standard
way, by the red sequence in galaxy clusters.}{}


\maketitle

\section{Introduction}
From an observational viewpoint, one of the fundamental question
of cosmology is measuring cosmological distances and then to build
up a suitable and reliable cosmic distance ladder. This issue has
recently become even more important due to the evident degeneracy of several dark
energy models with $\Lambda$CDM, despite the advent of the so-called
Precision cosmology, \citep{Ellis}.

More precisely, in the last two  decades, a class of accurate
standard candles, the Supernovae Ia (SNeIa) has been highly
studied and the results obtained from the use of these objects led
to the surprising discovery of the apparent acceleration of the
cosmic Hubble flow  (for a review see \citep{SCP}). However these
objects are hardly detectable at redshifts higher than $\sim 1.5$,
so we need distance indicators at higher redshifts in order to
remove the disturbing degeneration of dark energy models today
affecting the current cosmological picture ($\Lambda$CDM is a good
approximation of the observed Universe, also if there is yet no 
theoretical basis about the nature of its components, but the issue 
of global evolution is far from being addressed, for a comprehensive review see
\citep{copeland}). A possible way out for this problem could be
found by adopting Gamma Ray Bursts (GRBs) as distance indicators
also if it is premature, at the moment, to speak about standard
candles.

As it is well known, GRBs are  the most powerful explosions in the
Universe: the most likely scenarios for their generation are the
formation of  massive black holes or the  coalescence of binary
stellar systems. These events are observed at considerable
distances, so there are several efforts to frame them into the
standard of cosmological distance ladder. In literature, there are
several models that give account for the GRB formation. The
standard model \citep{Meszaros} predicts the formation of a black
hole originated by a massive star whose core is going to collapse.
Alternatively, the GRB phenomenon could be generated during an
accretion episode followed by a merging event which gives rise to
a jet-like outflow. For the class of short GRBs (which time
duration  is less than 2 seconds), the candidates are mergers of
neutron stars. Another model, \citep{Ruffini}, includes different
central energy sources and the formation of  charged black holes.

All  these scenarios retain essentially a similar shock
phenomenon: a "fireball" or a "fireshell". Here we will not go
into details, but it is worth stressing  that none of these models
is intrinsically capable of connecting all the observable
quantities.

Despite of the poor knowledge of the GRB mechanism, it seems that
GRBs could be used as reliable  distance indicators. In fact there
exist several observational correlations  among the photometric
and spectral properties of GRBs which point out that it could be
realistic to suppose them as distance indicators,
\citep{Leandros, Ghirl2}. Nevertheless the origin of
these spectroscopic and photometrical correlations is not known very well 
and there are several efforts to interpret
the behavior of GRB features in a coherent way, by  relatively
simple scenarios (e.g. see \citep{MG, ghisellini}). 
Succeeding in explain the mechanism that generates the GRBs is one of the 
objectives of the modern astrophysics and to clarify these observed 
correlations in this context would make GRBs as reliable distance indicators.
A complete review of the existing luminosity relations for GRBs can be found
in \citep{Schaefer}.

In this paper, we consider two relations, the one by
Liang-Zhang (LZ), \citep{LZ}, and the one by Ghirlanda (GGL),
\citep{GGL}. They are the only 3-parameters relations  and  have
less scatter with respect to the theoretical best fit than the
other 2-parameters ones. In a recent paper, \citep{CI}, starting
from a sample of GRBs,  a GRB-Hubble diagram has been derived
considering these relations. However it is worth noticing that the
calibration of the used relations has been necessary in order to
avoid the circularity problem. This means that all the relations
need to be calibrated for every set of cosmological parameters.
Indeed, all GRB distances, obtained in a photometric way, are
strictly dependent on the cosmological parameters since,
currently, there is no low-redshift ($z$ up to 0.2-0.3) set of
GRBs to achieve a cosmology-independent calibration. In order to
overcome this difficulty, Liang et al., \citep{Liang}, proposed a
method in which several GRB-relations have been calibrated by
SNeIa. In fact, supposing that our relations work at every
redshift and that, at the same redshift, GRBs and SNeIa have the
same luminosity distance, it becomes possible, in principle, to
calibrate the GRB-relations using an interpolation algorithm. In
this way,  it becomes possible building a GRB-Hubble diagram by
calculating the luminosity distance for each GRB with the
well-known relation between the luminosity distance $d_l$ and the
energy-flux ratio of the distance indicators, i.e.

\begin{equation}
\label{lum1}
 d_l = \left(\frac{E_{iso}}{4\pi S_{bolo}'}\right)^{\frac{1}{2}},
\end{equation}
where $E_{iso}$ is the  isotropic energy emitted in the burst and
$S_{bolo}'$ is the bolometric fluence corrected to the rest frame
of the source in consideration. This result can be connected to
the Hubble series, \citep{Visser1}, and the density parameters
$\Omega_M$ and $\Omega_{\Lambda}$ can be obtained. The  results in
\citep{CI} were in agreement with the other observations but the
estimation of the CPL-parameters \citep{CPL}, which is an
advantageous parameterization of cosmological Equation of State
(EoS) $w = p/\rho$, (see for a brief review of the various
parameterizations of the EoS the work by \citep{Alcaniz}), is only
marginally consistent with the data existing in the literature,
(e.g. see the already cited work by Visser). The reason of this
disagreement is due to the fact that the method used in \citep{CI}
works very well at redshift less than $z\simeq 1$ while CPL
parameters are supposed to work also at high redshift.

The aim of this paper is to take into account  a cosmological EoS
working at any redshift, using GRBs as tracers and adopting again
the CPL parameterization. The layout of the paper is the
following: in Sect. 2, we discuss the method which, in principle,
should allow to obtain a cosmology-independent formulation of the
luminosity distance and then of the distance modulus. Sect. 3 is
devoted to a discussion of the GRB luminosity-relations considered
in this work. In Sect. 4, we illustrate the fitting of the data
obtained by these relations while results and perspectives of the
approach are discussed in Sect. 5.

\section{The cosmological model}

The goal is to obtain an analytic formulation of the Hubble
diagram valid, in principle, at any redshift. Let us start from
the Friedmann equation

\begin{equation}\label{eq:nostart}
 H^2 = \frac{8 \pi G}{3} \rho - \frac{k c^2}{a^2}\,.
\end{equation}
 We obtain, by some algebra,  the following equation in terms of
 the density parameter

\begin{equation}
 H^2 = H_0^2 \left[\Omega_0\left(\frac{a_0}{a}\right)^{3(w+1)} -
 \left(\Omega_0 - 1 \right)\left(\frac{a_0}{a}\right)^2\right],
\end{equation}
where the subscript $0$ indicates the present value of the
parameters. From now onwards, we take into account a spatially
quasi-flat Universe, $k \approx 0$, the contribution of the
curvature will be negligible and we have $\Omega_0 \approx 1$, as
suggested by the latest CMBR \citep{WMAP} and  the SNeIa
observations \citep{SCP}. However in the final section, we will do
a test to verify this assumption with  observations coming from
GRBs. Now if we translate in terms of redshift $z$,

\begin{equation}
 \frac{a_0}{a} = 1+z\,,
\end{equation}
the previous equation reduces to

\begin{equation}\label{eq:no1}
 H^2(z) = H_0^2 \left( 1 + z\right)^{3(w+1)}\,.
\end{equation}
The $w$-parameter indicates  the EoS $w = p/\rho$, where $p$ and
$\rho$ are  the pressure and the matter-energy density of the
Universe, respectively. Considering the CPL parameterization of
the EoS, \citep{CPL}:

\begin{equation}
 w(z) = w_0 + w_a \frac{z}{1+z},
\end{equation}
and substituting  into Eq.(\ref{eq:no1}), we obtain:

\begin{equation}
 H(z) = H_0 \left[\left(1+z\right)^{\frac{3}{2}(w_0 + w_a + 1)} \exp{\left(\frac{-3w_a z}{2(1+z)}\right)}\right],
\end{equation}
which enters directly in the expression of the distance modulus

\begin{equation}\label{eq:no2}
\mu(z) = -5 + 5 \log{d_l(z)}\,,
\end{equation}
where $ d_l(z) = c (1+z) D_l(z)$ and where
\begin{equation}\label{eq:no2a}
D_l(z)=\int_0^z \frac{d \xi}{H(\xi)}\,. \end{equation} This means
that an analytic expression for $\mu$ can be achieved. The
integral $D_l$ in Eq.(\ref{eq:no2a}) can be solved giving a Gamma
function of the first kind \footnote[1]{In our  case, the variable
of the Gamma function, $z$, is always positive so that we have no
problem of discontinuity in applying the Gamma function in the
following calculations.}:

\begin{eqnarray}\label{eq:Luca}
 D_l(z) =  \left(\frac{3 w_a}{2}\right)^{-\frac{1+3 w_0+3 w_a}{2}}
\nonumber
\\
\left. \exp{\left(\frac{3 w_a}{2}\right)} \Gamma \left[ \frac{1+ 3 w_0 + 3 w_a}{2}, \frac{3 w_a}{2 (1+\xi)}\right]
\right|_{\xi = 0}^{\xi = z}.
\end{eqnarray}

Substituting  such an expression in the distance modulus, we
obtain a model for  data fitting which could work, in principle,
at any $z$. It is important to stress that the obtained expression
for the Hubble parameter $H(z)$ is independent of the density
parameters, $\Omega_M$ and $\Omega_{\Lambda}$, provided that their sum is equal to 1.

It is  worth noticing that we are using the CPL parameterization
not only for the dark energy component, but  for the total
energy-matter density of the Universe. This assumption works
because dark and baryonic matter are contributing with a null
pressure while the radiation component is negligible in matter-
and dark energy-dominated eras. Furthermore, the analytical
formulation which we are adopting for the luminosity distance is
assumed valid at any redshift $z$.

\section{GRBs luminosity relations}

As we said,  GRBs are the most powerful explosions in the
Universe, so they can be observed up to very far distances.
Theoretically, they can be observed up to a redshift of the order
$z \sim 10$. Some considerations on the spectrum of such sources
are due at this point. The electromagnetic emission spectrum of
GRBs ranges from radio up to gamma wavelengths, but the main bulk
of emission is in the gamma band. In the last years, thanks to
several spacecraft missions capable of observing this high energy
region, the main features of GRBs have been better known.
Recently, some photometric and spectroscopic relations between GRB
observables have been found and then the hypothesis that these
objects could be considered  suitable distance indicators has been
seriously considered. Nevertheless, up to now, there is no
theoretical model that fully explains these relations so the GRBs
cannot be considered as standard candles in a proper sense. For a
detailed review of the observational features see
\citep{Schaefer}.

Here, we are taking into account the existing 3-parameter
relations. This choice has been done because these relations put
the better constraints on the data giving less scatter between the
theoretical relation and the experimental data (e.g. see
\citep{Schaefer}). The first relation is the so-called Liang-Zhang
relation, \citep{LZ}, which allows to connect the GRB peak energy,
$E_p$, with the isotropic energy released in the burst, $E_{iso}$,
and with the  jet break - time of the afterglow optical light
curve in the rest frame, measured in days, $t_b$, that is
\begin{equation}\label{eq:noLZ}
 \log{E_{iso}}=a + b_1 \log{\frac{E_p (1+z)}{300keV}} + b_2 \log{\frac{t_b}{(1+z)1day}}
\end{equation}
where $a$ and  $b_i$, with $i=1,2$, are calibration constants.

The other one is the relation given by Ghirlanda et
al. \citep{GGL}. It connects the peak energy $E_p$ with the
collimation-corrected energy, or the energy release of a GRB jet,
$E_{\gamma}$, where

\begin{equation}
 E_{\gamma} = F_{beam} E_{iso}\,.
\end{equation}
and $F_{beam} = 1 - \cos{\theta}$, with $\theta_{jet}$ the jet
opening angle defined in \citep{Sari}:

\begin{equation}
 \theta_{jet} = 0.163\left(\frac{t_b}{1 + z}\right)^{3/8}\left(\frac{n_0\eta_{\gamma}}{E_{iso,52}}\right)^{1/8},
\end{equation}
where $E_{iso,52} = E_{iso}/10^{52}$ ergs,  $n_0$ is the
circumburst particle density in 1 cm$^{-3}$, and $\eta_{\gamma}$
the radiative efficiency. The Ghirlanda et al. relation is

\begin{equation}\label{eq:noGGL}
 \log{E_{\gamma}} = a + b \log{\frac{E_p}{300 keV}},
\end{equation}
where $a$ and $b$ are two calibration constants.

From these relations, we can obtain directly the luminosity
distance $d_l$ from the well-known formula which connects $d_l$
with the isotropic energy $E_{iso}$ and the bolometric fluence
$S_{bolo}$ :
\begin{equation}
 d_l = \left(\frac{E_{iso}}{4 \pi S_{bolo}}\right)^{\frac{1}{2}},
\end{equation}
from which it is easy to compute, for each GRB, the distance
modulus $\mu = $ and its error given by \citep{Liang}:
\begin{equation}
 \sigma_{\mu} =  \left[\left(2.5 \sigma_{\log{E_{iso}}}\right)^2  + \left(1.086\sigma_{S_{bolo}}/S_{bolo}\right)^2\right]^{\frac{1}{2}}
\end{equation}
with $\sigma_{\log{E_{iso}}}$ and $\sigma_{S_{bolo}}$  obtained
from the error propagation applied to Eq.(\ref{eq:noLZ}) and
Eq.(\ref{eq:noGGL}).  Moreover, we assume that the error in the
determination of the redshift $z$ is negligible, as well as for
the radiative efficiency $\eta_{\gamma}$. We note also that the
assumption of a well-known $n_0$ is a strong hypothesis since the
goodness of the fits depends, in particular, on this parameter.
However, we are going to consider the $n_0$ values for each GRB
given in the Table \ref{table:no2} also if it lacks a complete and
clear physical basis for the considered relations,
\citep{Friedman}. The  GRB data sample  is taken from the already
cited work by Schaefer. We  take into account 27 events with
extremely precise data. Such a sample is the same adopted in
\citep{CI}.

\section{The data fitting}

The next step is the  fit of the GRB sample with the empirical
relations, Eqs.(\ref{eq:noLZ}),(\ref{eq:noGGL}), described in
Sect. 3. The aim is to achieve an estimate of the CPL parameters
and consequently to determine the trend of the EoS at any
redshift, using the analytical relation, Eq.(\ref{eq:Luca}). As we
said,  we are considering the same sample of 27 GRBs used in
\citep{CI}, see Table \ref{table:no2}, in which we have added the
sample of SNeIa  by the Union Supernova Survey, \citep{SCP}.

\begin{center}
\begin{figure}
\includegraphics[width=9cm, height=6 cm]{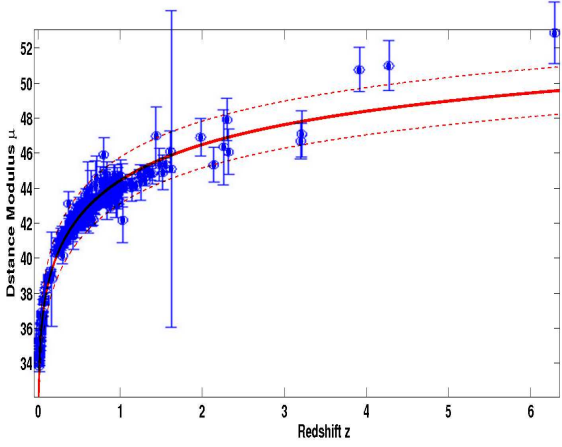}
\caption{Redshift-Distance modulus diagram for the GRB+SNeIa
sample. The black dots are the GRBs , the blue ones are the SNeIa.
The red line is the best fit obtained from the data, with the
dashed line representing the confidence limits at $3\sigma$. The
error bars on the Supernova data are not represented because they
are negligible.}
\label{fig:no1}
\end{figure}
\end{center}

\begin{figure}
\includegraphics[width=9 cm, height=6.5 cm]{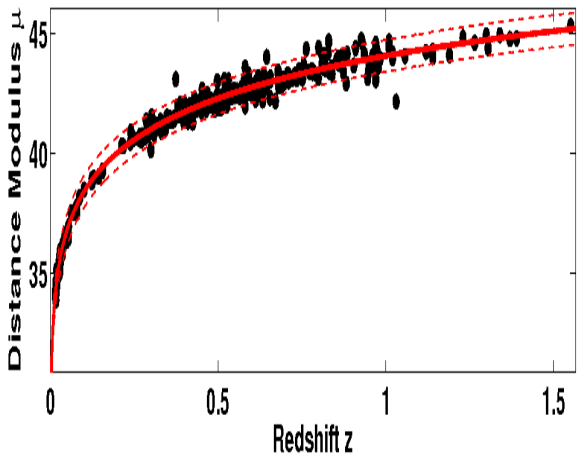}
\caption{Redshift-Distance modulus diagram for the SNeIa sample
only. The black dots are the Supernova data while the red line is
the best fit obtained from the data. The error bars on the
Supernova data are not represented because they are negligible.}
\label{fig:no2}
\end{figure}

The numerical results of the fits are shown in Table
\ref{table:no1}, where we obtain a robust estimation of the
CPL parameters for both the relations used, with and without SNeIa
data.  An immediate comparison is done with the best fit applied
only to the SNeIa sample. It is evident how adding GRBs to SNeIa
data improves the knowledge and the precision on the EoS parameter
$w$. In Fig. ~\ref{fig:no1}, the best fit curve, in the case of LZ
relation, is plotted  while, in Fig. ~\ref{fig:no2}, it is plotted
the best fit curve for the SNeIa data using the theoretical model
described previously.
\begin{table}
\caption{Results of the fits. SNeIa is only for the Supernova Ia data,
LZ is for the GRBs data obtained from the Liang-Zhang relation,
GGL for the Ghirlanda et al. one. Note the improvement on the $w$-parameter
using the GRBs data in addition to the  SNeIa data corrected for the 3 ``outlier'' GRBs.} 
\label{table:no1} 
\centering 
\begin{tabular}{l c c c} 
\hline\hline 
Relation & $w_0$ & $w_a$ & $R^2$ \\ 
\hline 
 SNeIa & $-0.910 \pm 0.070$ & $0.755 \pm 0.054$ & $0.983$ \\
 LZ & $-1.39 \pm 0.38$ & $1.18 \pm 0.37$ & $0.817$ \\
 GGL & $-1.46 \pm 0.38$ & $1.36 \pm 0.32$ & $0.812$ \\
 LZ + SNeIa & $-1.15 \pm 0.10$ & $0.93 \pm 0.11$ & $0.933$ \\
 GGL + SNeIa & $-1.42 \pm 0.12$ & $1.24 \pm 0.13$ & $0.920$ \\
\hline 
\end{tabular}
\end{table}

In order to measure the goodness of the fit, we use the $R^2$
test for an accurate reliability, see Table \ref{table:no1}. The
$R^2$ test is a measure of how successful the fit is in
explaining the variation of the data (see for details
\citep{Draper}). Exactly an $R^2$ close to 1.0 indicates that we have accounted for almost all of the variability with the data specified in the model. As standard, the $R^2$ test is the square of
the correlation between the response values and the predicted
response values, that is:
\begin{equation}
 R^2 = 1 - \frac{SSE}{SST} = 1 - \frac{\sum_{i = 1}^{n} w_i (y_i - \hat{y}_i)}{\sum_{i = 1}^{n} w_i (y_i - \bar{y}_i)^2},
\end{equation}
where $SSE$  is the sum of the squares due to errors and it
measures the total deviation of the response values from the fit and
 $SST$ is the sum of squares about the mean: $\hat{y}$ is the predicted response value, $\bar{y}$ is the
mean value and the $w_i$ are the weights on the values. 

\begin{center}
\begin{figure}
\includegraphics[width=8 cm, height=6 cm]{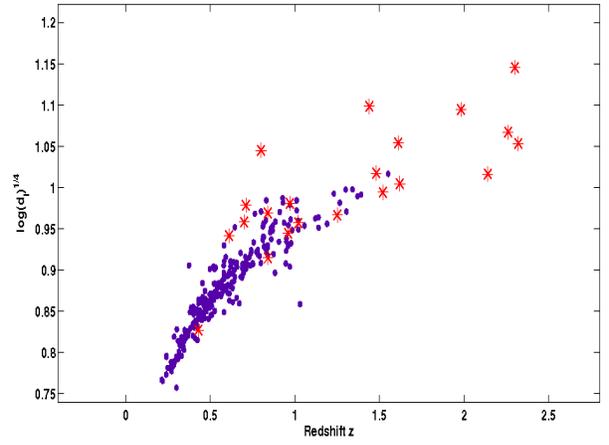}
\caption{Comparison between GRBs-SNeIa data using the relation
$(\log{d_l})^{1/4}$ for each data sample. Blue dots are the SNeIa
while the red star are the GRBs. We are considering  redshifts
between 0 and 3, in order to  better analyze the data trend in the
overlapping redshift range.}
\label{fig:no4}
\end{figure}
\end{center}

A further consistency test for the adopted samples of GRBs and
SNeIa can be derived considering the relation
$(\log{d_l})^{1/4}-redshift$ vs $z$.  In such a way, data from GRBs
and SNeIa can be better separated  in order to track the whole
trend of both sets and showing, in particular, the spreading of
GRB distribution.  In Fig. \ref{fig:no4}, the results of this
relation is plotted and it is reasonable to conclude that SNeIa
could be used to calibrate GRBs at lower redshifts. In particular,
it is evident, even considering the only   SNeIa data,  a change
in the trend at redshift included between $z = 0.3$ and $z = 0.5$.

\begin{center}
\begin{figure}
\includegraphics[width=9cm, height=6.5 cm]{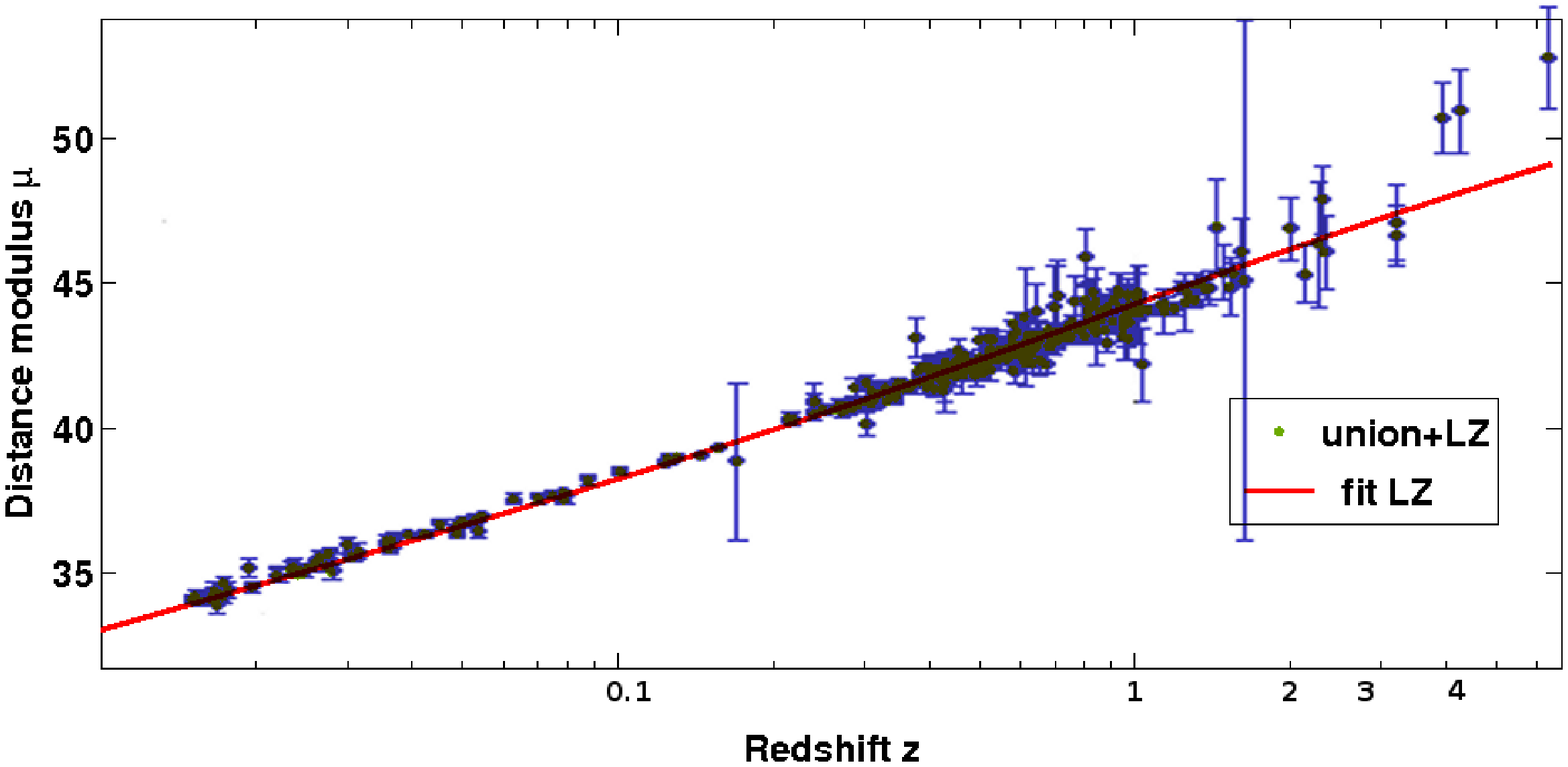}
\caption{Redshift-Distance modulus diagram for the GRB+SNeIa
sample versus redshift in logarithmic scale. }
\label{fig:no5}
\end{figure}
\end{center}

\begin{figure*}
\centering
\begin{tabular}{|c|c|}
\hline
\includegraphics[scale=0.6]{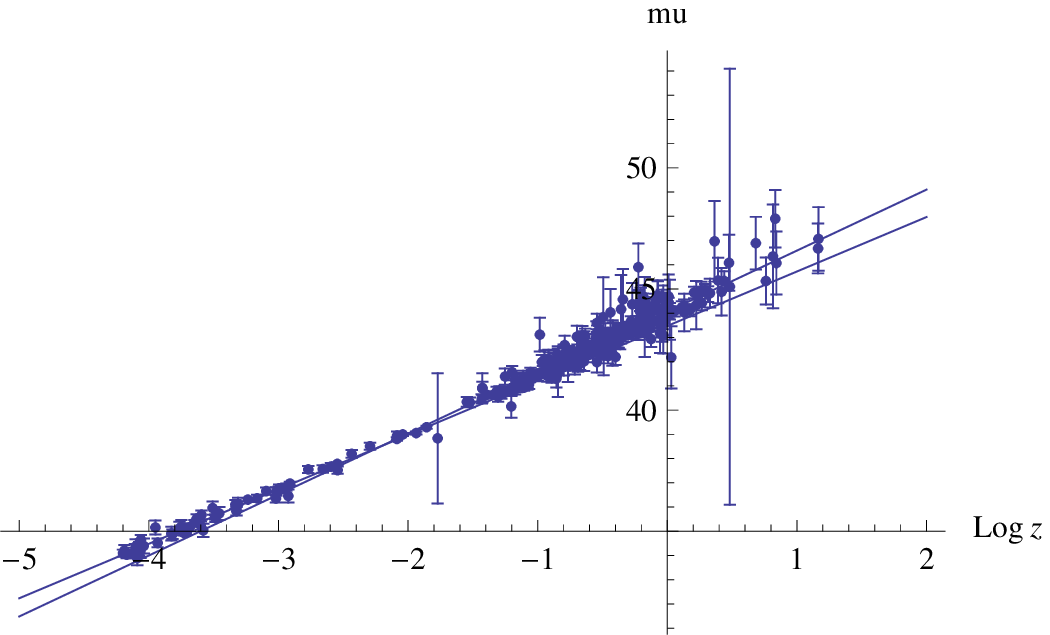}&
\includegraphics[scale=0.6]{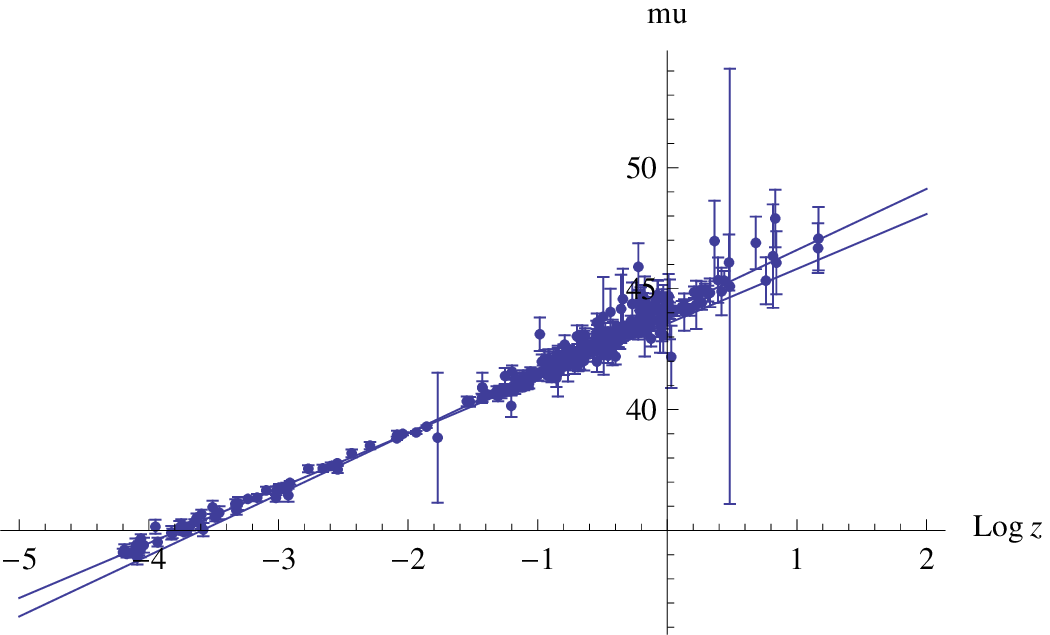}\tabularnewline
\hline
\includegraphics[scale=0.6]{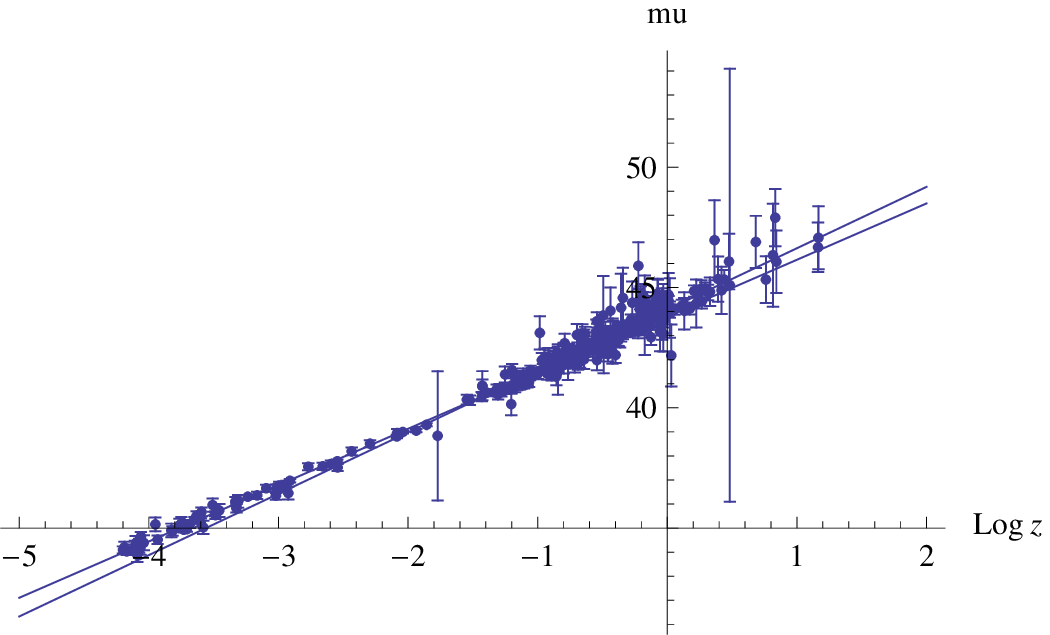}&
\includegraphics[scale=0.6]{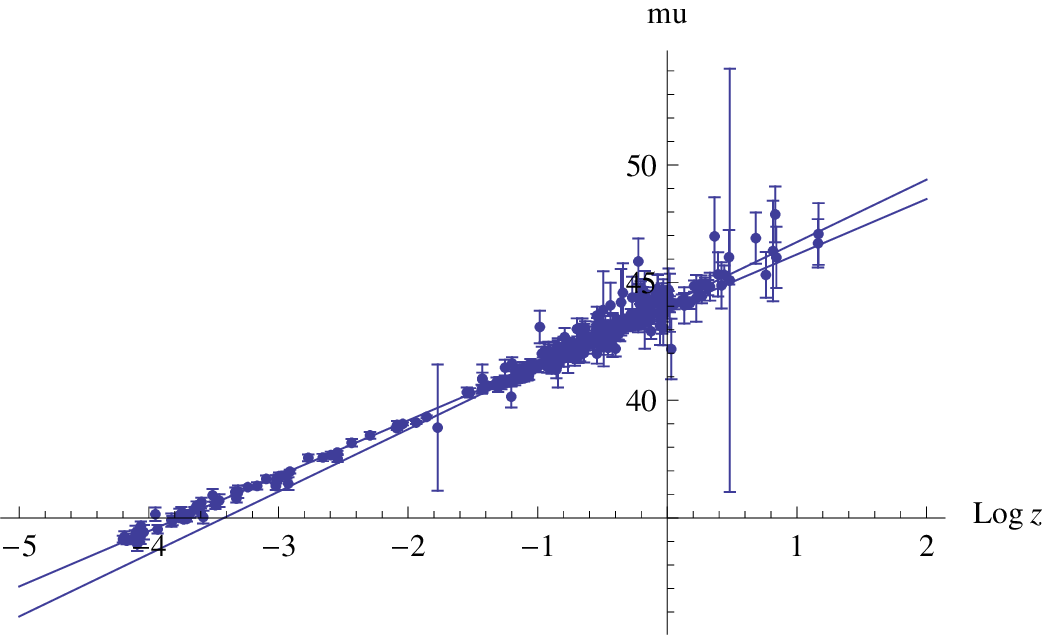}\tabularnewline
\hline
\includegraphics[scale=0.6]{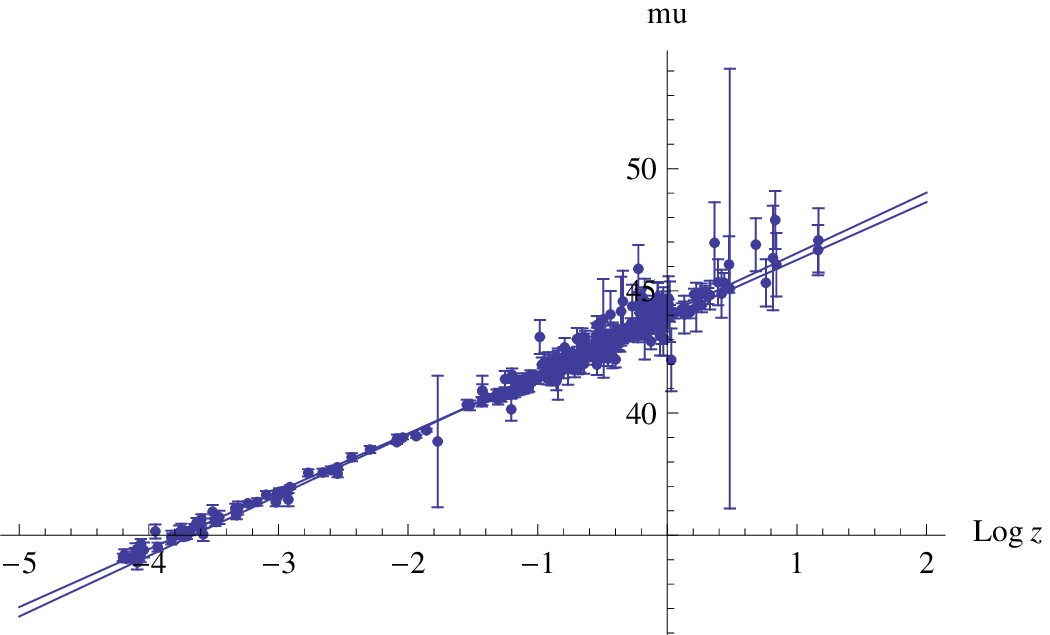}&
\includegraphics[scale=0.6]{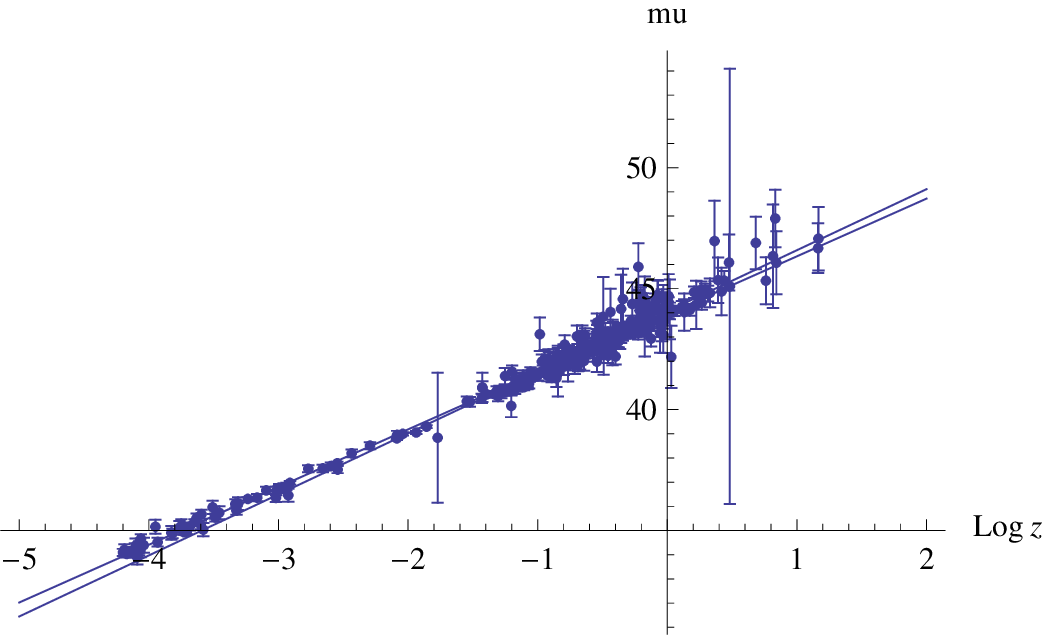}\tabularnewline
\hline
\end{tabular}
\caption {Redshift-Distance modulus diagram for the GRB+SNeIa
samples versus $\log{z}$ as described in the text. Note the ''linearization`` of the two best
fit curves of the two data samples, done with the model $a_i + b_i \log{z}$, obtained considering only GRB and Sn
data up to and beyond, respectively, the cut-redshift value $z_t$, with $z_t$ ranging from $z_{tmin} = 0.1$ and $z_{tmax} = 0.6$ with step of $z_{step} = 0.1$.}
\label{fig:no6}
\end{figure*}

In particular, the extension of the Supernova Hubble Diagram with the GRB data can
be used to improve our knowledge of the trend at high redshift. In this way, using
also the GRB data, we build a plot, Fig. ~\ref{fig:no5}, where the
distance modulus $\mu$ versus the redshift $z$, in a logarithmic
scale, is plotted. The best fit curve, obtained with the  method
described previously, Eq.(\ref{eq:Luca}), is also reported.
However in order to better point out  this variation, we have done
an extended analysis, Fig. ~\ref{fig:no6}, where we considered the GRB+Sn sample up to a certain value of the redshift $z_t$ and its complementary,
and then we fitted these samples with a simple curve of the type $a_i + b_i \log{z}$, with the cut-redshift $z_t$ ranging from $z_t = 0.1$ to
$z_t = 0.6$ with redshift-step $0.1$. The result of this analysis is given in Table \ref{table:no5},
where we note that at redshift greater than $z_t = 0.5$, the two best fit
curves tend slowly to coincide, suggesting that something happens at
redshift smaller than this value. This result led in the past to the
conclusion that the universe is accelerating.

\begin{figure}
\includegraphics[width=8.5 cm, height=6 cm]{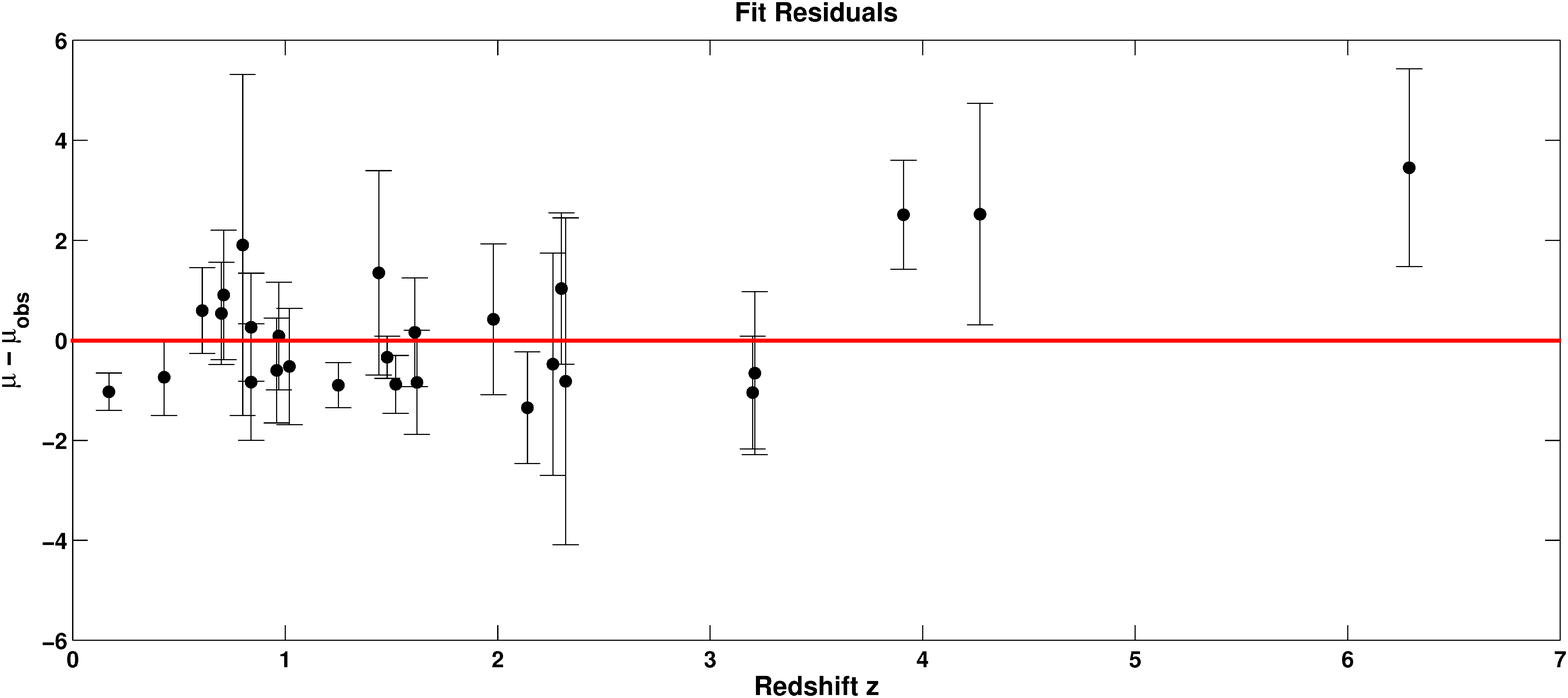}
\caption{Comparison between  the best fit of $\mu$ and the
observed distance modulus $\mu_{obs}$ at any redshift. The black
dots are the GRBs data and the red line is the best fit curve
representing the theoretical distance modulus.} 
\label{fig:no3}
\end{figure}

\begin{table*}[ht]
\caption{Results of the logarithmic fit described in the text and plotted in fig. \ref{fig:no6}}. 
\label{table:no5} 
\centering 
\begin{tabular}{c c c c c} 
\hline\hline 
$z_t$ & $a_1$ & $b_1$ & $a_2$ & $b_2$ \\ 
\hline 
 $0.1$ & $44.07$ & $2.52$ & $43.48$ & $2.25$  \\
 $0.2$ & $44.07$ & $2.52$ & $43.55$ & $2.28$  \\ 
 $0.3$ & $44.08$ & $2.55$ & $43.81$ & $2.34$  \\
 $0.4$ & $44.07$ & $2.65$ & $43.85$ & $2.35$  \\ 
 $0.5$ & $44.06$ & $2.48$ & $43.90$ & $2.37$  \\
 $0.6$ & $44.06$ & $2.52$ & $43.95$ & $2.38$  \\ 
\hline 
\end{tabular}
\end{table*}

It is suggestive to link the observed behaviour
of the comsological accelletration to the 
the formation and evolution of astrophysical structures.
For instance, the infalling rate of
field galaxies \citep{Stott} could be related to the overal expansion rate.

It would be very interesting to asses quantitatively such a relation,
from the massive galaxy clusters with some good
indicator at that redshift in order to clarify a possible
correlation between these two phenomena. All of these cosmological
and astrophysical hints lead us to reconsider from a different
point of view the possible cause of the apparent acceleration of
the Universe which could be addressed to some more astrophysical standard
effects, (Izzo et al. in preparation).

In Fig. ~\ref{fig:no3}, it is plotted the comparison between the
theoretical $\mu_{th}$ and the observed distance modulus
$\mu_{obs}$ at any redshift, the residual plot. A smooth trend up
to $ z \approx 3.5$ in the residual curve can be immediately
detected.  Beyond this limit, we have 3 GRBs that exceed, by the
same side, the $3\sigma$ confidence limit of the best fit. This
discrepancy is clear in Fig. ~\ref{fig:no5}, where  is plotted the
best fit for the combined sample in the case of LZ relation with a
logarithmic scale for the redshift. It is worth noticing that a
similar, but opposite discrepancy was obtained by
\citep{Leandros2} using only the SNeIa sample. In that paper,  the
authors found a brightening for the SNeIa at high redshift, while
here, we find a sort of ``darkening`` for the GRBs at redshift
higher than $z = 4$.

This fact is fundamental  for the goodness of the fit because
these GRBs represent the most distant objects that one can use to
make such an analysis and their weight on the fit is very high, in the sense that they appear to be not accurate 
 distance indicators.
There could be several explanation for this anomalous GRB
brightness at high redshift and the most likely are the following:
\begin{itemize}
 \item there is some process of absorption of gamma radiation where the GRB $\gamma$
 photons may interact with the very low energy photons incoming from
 the cosmic thermal background radiation, \citep{Zdziarski};
 \item these 3 GRBs could be outliers;
 \item the Circum Burst Density (CBD) in the host galaxy at that epoch, $z > 4$,
 could be different from the CBD in galaxies at low redshift;
 \item there could be some high-energy bias: since at that distances only
 very powerful GRBs can be observed, some high-energy process,
 involving very energetic $\gamma$-photons, \citep{Kelner, Razzaque},
 could happen so that the flux received by our detectors is
 dimmed;
 \item it could be that the CPL parameterization, or the $\Lambda$CDM model,
 is a bad approximation for the cosmological EoS.
\end{itemize}

It is very interesting to note that this phenomenon is very similar to the change in the trend of the Supernovae Ia data, as we mentioned earlier. 
However, at these redshifts is very difficult to make any kind of astrophysical constraint, since there is not enough information about this region of the Universe.

\begin{center}
\begin{figure}
\includegraphics[width=9 cm, height=6.5 cm]{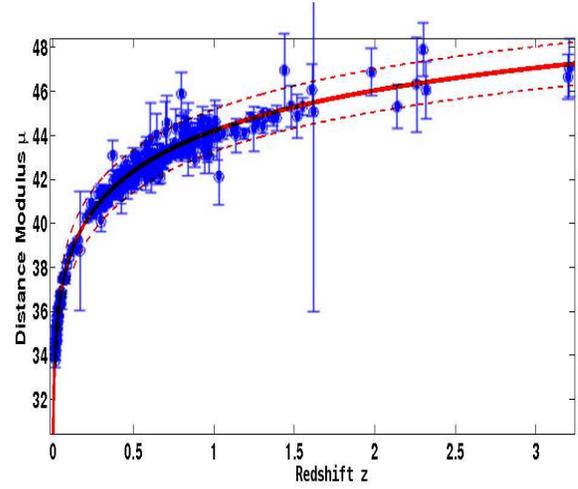}
\caption{Redshift-Distance modulus diagram for the corrected GRB+SNeIa
sample. The red line is the best fit obtained from the data, with the
dashed line representing the confidence bounding at $3\sigma$.}
\label{fig:no7}
\end{figure}
\end{center}

\begin{center}
\begin{figure}[ht]
\includegraphics[width=8.5cm, height=6.5 cm]{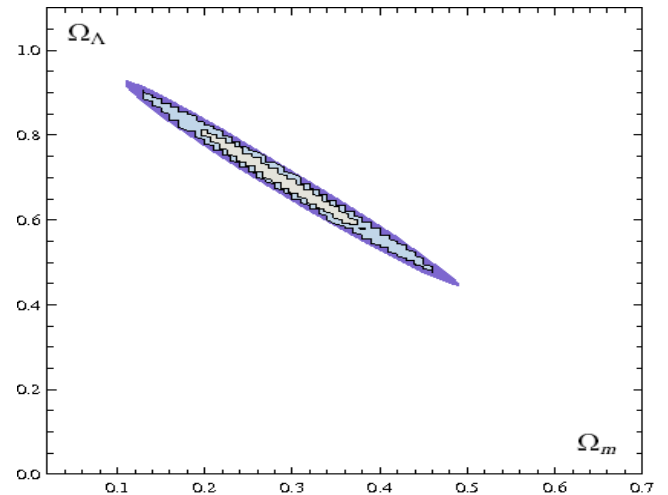}
\caption{68\%, 95\% and 98\% constraints on $\Omega_m$ and
$\Omega_{\Lambda}$, see Fig.( ~\ref{fig:no8}) obtained from UNION
sample and the GRB sample corrected for the 3 wrong GRBs. }
\label{fig:no8}
\end{figure}
\end{center}

Nevertheless, we repeat the analysis described above without these 3
GRBs obtaining a better value than the previous one for the
$R^2$ test. The results of these corrected fits are shown in
the table \ref{table:no3}. In Fig. ~\ref{fig:no7}, it is plotted
the best fit with this corrected sample. From these results we
conclude that the complete sample gives different results from the
corrected sample, the first one suggesting a phantom/quintessence
regime for the present epoch while the second one can be enclosed
in the case of an accelerating $\Lambda$CDM model. This last
result is confirmed by the following analysis, where we have
performed a Monte-Carlo-like procedure for the comparison of the
results with the usual likelihood estimator given by
\begin{equation}
\chi^2 = \sum_{i = 1}^{N} \left[\frac{(\mu_{th}(z_i) - \mu_{obs}(z_i))^2}{\sigma_i}\right],
\end{equation}
in the context of a $\Lambda$CDM model of the Universe,  where
$\mu_{th}$ is the distance modulus computed from the
Eq.(\ref{eq:no2}) and Eq.(\ref{eq:no2a}), $z_i$ is the  observed
redshift for each GRB and $\sigma_i$ the observed distance modulus
uncertainty. The results of this analysis are shown in the Table
\ref{table:no4}, where we can see the improvement obtained by the
GRB sample corrected for the 3 "wrong" GRBs. In
Fig.(\ref{fig:no8}), it is shown the contour plot of the corrected
sample, where the boundaries correspond to 1$\sigma$, 2$\sigma$
and 3$\sigma$ confidence levels. In this, case we do not consider
a purely flat geometry, as we previously said in the Sect. 2, but
we constraint the $k$-parameter of Eq.(\ref{eq:nostart}) to vary
between the value $-0.05<k<0.05$ so that we take account of a
possible small contribution due to the curvature density.

\begin{center}
\begin{figure}[ht]
\includegraphics[width=8.5cm, height=6.5 cm]{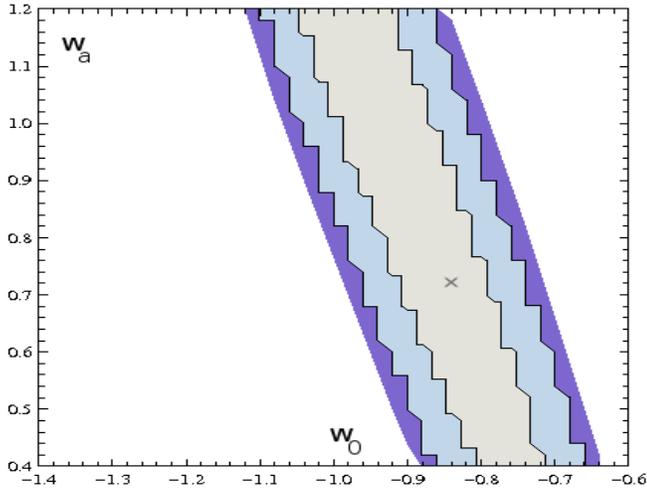}
\caption{68\%, 95\% and 98\% constraints on $w_0$  and $w_a$
obtained from UNION sample and the GRB sample corrected for the 3
wrong GRBs. The cross represents the best fit value and it is in a
good agreement with what found using the theoretical model
described in Sect.2.} \label{fig:no9}
\end{figure}
\end{center}

We have adopted a similar procedure in the case of an EoS evolving
with redshift and where $\mu_{th}$ is obtained by the
Eq.(\ref{eq:Luca}). The result of this analysis is plotted in Fig.
~\ref{fig:no9} where the best fit value, the cross in the figure,
corresponds to the value $w_0 = -0.84 \pm 0.14$ and $w_a = 0.72
\pm 0.06$, in a good agreement with the results obtained, see
Table \ref{table:no3}, using our theoretical relation,
Eq.(\ref{eq:Luca}).

Summarizing, from this analysis, we conclude that the corrected
sample agrees fairly well with the $\Lambda$CDM model with a mall
contribution of the curvature parameter, being equal to $k = 0.01
\pm 0.04$. In other words, the method delineated in the Sect.2
seems a good approximation of the observed cosmography and agrees
very well with  the $\Lambda$CDM model so that we can argue  that
GRBs could be good distance indicators at redshift values up to $z
= 4$.

\begin{table*}[ht]
\caption{Cosmological density parameters, with uncertainties
computed at 1$\sigma$ confidence limit, obtained from the MC-like procedure} 
\label{table:no4} 
\centering 
\begin{tabular}{l c c c c} 
\hline\hline 
Sample & $\Omega_m$ & $\Omega_{\Lambda}$ & $\Omega_k$ & $\chi^2$ \\ 
\hline 
 UNION + GRB & $0.26 \pm 0.14$ & $0.73 \pm 0.14$ & $0.01 \pm 0.04$ & $1.032$ \\
 UNION + GRB corrected & $0.25 \pm 0.10$ & $0.74 \pm 0.135 $ & $0.01 \pm 0.035$ & $1.00027$ \\ 

\hline 
\end{tabular}
\end{table*}

\section{Discussion and Conclusions}

Starting from the Friedmann equation, we have investigated a new
method to constrain the cosmological Equation of State at high
redshifts.  The working hypothesis lies on the use of GRBs as
distance indicators  at  high redshift, well beyond the distance
where SNeIa are detected up to date.  The CPL parameterization for
the EoS has been explicitly used for the whole matter-energy
content of the Universe as a suitable approach to investigate the
parameter $w=w(z)$ and discriminate with respect to the
$\Lambda$CDM model. In particular, regarding the Friedmann
equations, for which

\begin{equation}
\frac{H'}{H} = \frac{(1 + q)}{(1+z)},
\end{equation}

where $q$ is the deceleration parameter and  where  the prime
denotes the derivative with respect to the redshift, we have
obtained, in the case of the LZ relation, with a reliable
confidence level, the epoch for the transition between the
deceleration-acceleration phases at a redshift value of $z \approx
5 $. This is a value that, also if higher than the redshift of the
farther GRB used, could be in agreement with current
quasar formation scenarios. Besides, we are in good agreement with
the observed phantom/quintessence regime at present epoch, that is
for $z \rightarrow 0$, we obtain $w \leq -1$. However we have
found an anomaly at $z \approx 4$ and beyond for which we rejected
3 GRBs at that distances. Several explanations are possible for
this problem which can be summarized as:
\begin{itemize}
 \item the effective cosmology is not described by the $\Lambda$CDM model
 \item the GRBs evolve with the redshift
 \item there is some nuclear or
electromagnetic processes between the $\gamma$ photons and the
baryons involved in a  GRB that would dim the observed flux or
fluence.
\end{itemize}

Nevertheless such issues  still remain unsolved and it represents
a  challenge. With this fact in mind,  we have performed  the same
analysis without these 3 GRBs obtaining different results from the
previous ones. In particular,  we rejected the today phantom
regime by this new analysis, obtaining for $w_0$ a value  in
agreement with the $\Lambda$CDM model at present epoch. The
method, also if preliminary, seems to indicate that GRBs could be
actually used as standard candles once a reliable unified model
for the photometric and spectroscopic quantities is achieved
(reliable results in this sense are presented in
\citep{ghisellini}). However, more robust samples of data are
needed and more realistic EoS  (with respect to the simple perfect
fluid models) should be taken into account in order to suitably
track  redshift at any epoch (see for example \citep{elizalde}).

With the improving of the observations, in particular with the
launch of new satellites devoted to the GRB surveys, as
Fermi-GLAST\footnote[3]{http://fermi.gsfc.nasa.gov} and
AGILE\footnote[4]{http://agile.rm.iasf.cnr.it}, one should be able
to expand the samples of GRBs, possibly with data coming from
objects at higher redshift.

In summary, considering  these preliminary results, it seems that
GRBs could be considered as a useful tool to remove degeneration
and constrain self-consistent cosmological models. Furthermore the
matching with other distance indicators  would improve the
consistency of the Hubble distance-redshift  diagram by extending
it up to redshift $6\, -\, 7$ and over.

\hspace{1cm}

We wish to thank Riccardo Benini for useful discussions and
suggestions.

\begin{table}
\caption{Results of the fits corrected for the 3 ``wrong'' GRBs.
SNeIa is just for the Supernova Ia data, LZ is for the GRBs data obtained from the Liang-Zhang relation,
GGL for the Ghirlanda et al. one.} 
\label{table:no3} 
\centering 
\begin{tabular}{l c c c} 
\hline\hline 
Relation & $w_0$ & $w_a$ & $R^2$ \\ 
\hline 
 LZ + SNeIa & $-0.95 \pm 0.01$ & $0.74 \pm 0.01$ & $0.999$ \\
 GGL + SNeIa & $-0.865 \pm 0.005$ & $0.66 \pm 0.005$ & $0.999$ \\
\hline 
\end{tabular}
\end{table}

\begin{table*}
\centering 
\caption{GRBs Data Sample} 
\label{table:no2} 
\begin{tabular}{c c c c c c c}
\hline\hline 
$GRB$ & $z$ & $E_p$ (keV) & $S_{bolo}$ (erg cm$^{-2}$) & $t_{jet}$ (days) & $\theta_{jet}$ (deg.) & $n_0$ $(cm^{-3})$ \\
(1) & (2) & (3) & (4) & (5) & (6) & (7)\\
\hline 
970508  & 0.84 & 389 $\pm$ 40 & 8.09E-6 $\pm$ 8.1E-7    &   25  $\pm$   5 & 23  $\pm$   3 &  3.0 $\pm$ 2.4 \\ 
970828  & 0.96 & 298 $\pm$ 30 & 1.23E-4 $\pm$   1.2E-5  &   2.2 $\pm$   0.4 &   5.91    $\pm$   0.79 & 3.0 $\pm$ 2.4  \\
980703  & 0.97 & 254 $\pm$ 25 & 2.83E-5 $\pm$   2.9E-6  &   3.4 $\pm$   0.5 &   11.02   $\pm$   0.8 & 28.0 $\pm$ 10   \\
990123  & 1.61 & 604 $\pm$ 60 & 3.11E-4 $\pm$   3.1E-5  &   2.04    $\pm$   0.46    &   3.98    $\pm$   0.57 &  3.0 $\pm$ 2.4 \\
990510  & 1.62 & 126 $\pm$ 10 & 2.85E-5 $\pm$   2.9E-6  &   1.6 $\pm$   0.2 &   3.74    $\pm$   0.28 &  0.29 $\pm$ 0.14 \\
990705  & 0.84 & 189 $\pm$ 15 & 1.34E-4 $\pm$   1.5E-5  &   1   $\pm$   0.2 &   4.78    $\pm$   0.66 &  3.0 $\pm$ 2.4 \\
990712  & 0.43 & 65 $\pm$ 10 & 1.19E-5  $\pm$   6.2E-7  &   1.6 $\pm$   0.2 &   9.47    $\pm$   1.2 & 3.0 $\pm$ 2.4  \\
991216  & 1.02 & 318 $\pm$ 30 & 2.48E-4 $\pm$   2.5E-5  &   1.2 $\pm$   0.4 &   4.44    $\pm$   0.7 & 4.7 $\pm$ 2.8  \\
010222  & 1.48 & 309 $\pm$ 12 & 2.45E-4 $\pm$   9.1E-6  &   0.93    $\pm$   0.1 &   3.03    $\pm$   0.14 &  3.0 $\pm$ 2.4 \\
011211  & 2.14 & 59 $\pm$ 8 & 9.20E-6   $\pm$   9.5E-7  &   1.56    $\pm$   0.16    &   5.38    $\pm$   0.66 &  3.0 $\pm$ 2.4 \\
020124  & 3.20 & 87 $\pm$ 18 & 1.14E-5  $\pm$   1.1E-6  &   3   $\pm$   0.4 &   5.07    $\pm$   0.64 &  3.0 $\pm$ 2.4 \\
020405  & 0.70 & 364 $\pm$ 90 & 1.10E-4 $\pm$   2.1E-6  &   1.67    $\pm$   0.52    &   6.27    $\pm$   1.03 &  3.0 $\pm$ 2.4 \\
020813  & 1.25 & 142 $\pm$ 14 & 1.59E-4 $\pm$   2.9E-6  &   0.43    $\pm$   0.06    &   2.8 $\pm$   0.36 &  3.0 $\pm$ 2.4 \\
021004  & 2.32 & 80 $\pm$ 53 & 3.61E-6  $\pm$   8.6E-7  &   4.74    $\pm$   0.5 &   8.47    $\pm$   1.06 & 30.0 $\pm$ 27.0  \\
030226  & 1.98 & 97 $\pm$ 27 & 8.33E-6  $\pm$   9.8E-7  &   1.04    $\pm$   0.12    &   4.71    $\pm$   0.58 &  3.0 $\pm$ 2.4 \\
030328  & 1.52 & 126 $\pm$ 14 & 6.14E-5 $\pm$   2.4E-6  &   0.8 $\pm$   0.1 &   3.58    $\pm$   0.45 &  3.0 $\pm$ 2.4 \\
030329  & 0.17 & 67.9 $\pm$ 2.3 & 2.31E-4   $\pm$   2.0E-6  &   0.5 $\pm$   0.1 &   5.69    $\pm$   0.5 &  1.0 $\pm$ 0.11 \\
030429  & 2.66 & 35 $\pm$ 12 & 1.13E-6  $\pm$   1.9E-7  &   1.77    $\pm$   1.0 &   6.3 $\pm$   1.52 &  3.0 $\pm$ 2.4 \\
041006  & 0.71 & 63 $\pm$ 12 & 1.75E-5  $\pm$   1.8E-6  &   0.16    $\pm$   0.04    &   2.79    $\pm$   0.41 & 3.0 $\pm$ 2.4  \\
050318  & 1.44 & 47 $\pm$ 15 & 3.46E-6  $\pm$   3.5E-7  &   0.21    $\pm$   0.07    &   3.65    $\pm$   0.5 & 3.0 $\pm$ 2.4  \\
050505  & 4.27 & 70 $\pm$ 23 & 6.20E-6  $\pm$   8.5E-7  &   0.21    $\pm$   0.04    &   3.0 $\pm$   0.8 & 3.0 $\pm$ 2.4  \\
050525  & 0.61 & 81.2 $\pm$ 1.4 & 2.59E-5   $\pm$   1.3E-6  &   0.28    $\pm$   0.12    &   4.04    $\pm$   0.8 & 3.0 $\pm$ 2.4  \\
050904  & 6.29 & 436 $\pm$ 200 & 2.0E-5 $\pm$   2E-6    &   2.6 $\pm$   1   &   8   $\pm$   1 &  3.0 $\pm$ 2.4 \\
051022  & 0.80 & 510 $\pm$ 22 & 3.40E-4 $\pm$   1.2E-5  &   2.9 $\pm$   0.2 &   4.4 $\pm$   0.1 & 3.0 $\pm$ 2.4  \\
060124  & 2.30 & 237 $\pm$ 76 & 3.37E-5 $\pm$   3.4E-6  &   1.2 $\pm$       &   3.72    $\pm$   0.15 & 3.0 $\pm$ 2.4  \\
060210  & 3.91 & 149 $\pm$ 35 & 1.94E-5 $\pm$   1.2E-6  &   0.33    $\pm$   0.08    &   1.9 $\pm$   0.17 &  3.0 $\pm$ 2.4 \\
060526  & 3.21 & 25 $\pm$ 5 & 1.17E-6   $\pm$   1.7E-7  &   1.27    $\pm$   0.35    &   4.7 $\pm$   1 & 3.0 $\pm$ 2.4  \\
\hline 
\end{tabular}
\\
\hspace{1mm}
References: \citep{Jimenez}; \citep{Metzger}; \citep{Djorgovski}; \citep{Kulkarni}; \citep{Israel}; \citep{Bjornsson}; \citep{Li}
\end{table*}

\end{document}